\documentclass[12pt]{article} 
\usepackage{epstopdf}
\usepackage{tikz,pgf}

\newtheorem{theorem}{Theorem}

\newtheorem{definition}[theorem]{Definition}

\usepackage{amssymb}
\usepackage{amsmath}
\usepackage{graphicx}

\usepackage{hyperref}
\usepackage{epsf}
\usepackage{psfrag}
\hypersetup{colorlinks,linkcolor={red},citecolor={red},urlcolor={blue}} 

\newcommand{\key}[1]{\textbf{#1}}
\newcommand{\tsc}[1]{\textrm{\textsc{#1}}}

\newcommand{\alge}{%
\begin{tabbing}%
99 \= xxx\=xxx\=xxx\=xxx\=xxx\=xxx\=xxx\=xxx\=xxx\=xxx\=xxx \+ \kill 
}

\newcommand{\algeN}{%
\begin{tabbing}%
\hspace*{1pt}999\=xxx\=xxx\=xxx\=xxx\=xxx\=xxx\=xxx\=xxx\=xxx\=xxx\=xxx \+ \kill  
}

\newcommand{\algv}{%
\end{tabbing}
}

\newenvironment{algN}[1]{%
\vspace{4mm}  
\vbox\bgroup\noindent\tsc{#1}%
\vspace*{-2mm}  
\algeN}
{
\algv\egroup
\vspace{0mm}  
}

\begin{document}
\title{\bf Warshall's algorithm---survey and applications}
\maketitle

\pagestyle{myheadings}
\markright{Warshall's algorithm---survey and applications}

\begin{center}
{\bf Zolt\'an K\'ASA}

Sapientia Hungarian University of Transylvania, Cluj-Napoca

\href{http://www.ms.sapientia.ro/en}{Department of Mathematics and Informatics,} \\Tg. Mure\c s, Romania

\href{mailto:kasa@ms.sapientia.ro}{kasa@ms.sapientia.ro}
\end{center}

\begin{abstract} 
\noindent The survey presents the well-known Warshall's algorithm, a generalization and some interesting applications of this.
\end{abstract}

\section{Introduction} 

Let $R$ be a binary relation on the set $S=\{s_1, s_2, \ldots, s_n\}$, we write $s_iRs_j$ if $s_i$ is in relation to $s_j$.  The relation $R$ can be represented by the so called \textit{relation matrix}, which is $$A=(a_{ij}){\footnotesize_{
	\begin{array}{l} 
	i=\overline{1,n} \\ 
	j=\overline{1,n} \end{array}}
}, {\rm \  where } \ \  a_{ij}= \left\{
\begin{array}{ll} 
1 &  {\rm if \ } s_i R s_j\\ 
0 & {\rm otherwise.}\\
\end{array} \right.$$

\noindent The transitive closure of the relation $R$ is the binary relation $R^*$  defined as:

\noindent
$s_i R^* s_j$ if and only if there exists $s_{p_1}$, $s_{p_2}$, $\ldots$, $s_{p_r}, r\ge 2$ such that $s_i=s_{p_1} $, $s_{p_1}R s_{p_2}$, $s_{p_2} R s_{p_3}, \ldots, $ $s_{p_{r-1}} R s_{p_r},$ 
$s_{p_r}=s_j$. The relation matrix of $R^*$ is $A^*=(a^*_{ij})$.

\medskip\noindent Let us define the following operations. If $a,b\in \{0,1\}$ then $a+b=0$ for $a=0, b=0$, and $a+b=1$ otherwise. $a\cdot b=1$ for $a=1, b=1$, and $a\cdot b=0$ otherwise. In this case 

\[A^*=A+A^2+ \cdots + A^n.\]

\medskip\noindent The transitive closure of a relation can be computed easily by the Warshall's algorithm \cite{warshall}, \cite{baase}:

\begin{algN}{Warshall($A,n$)} 
	\hspace*{-20pt} \textit{Input:} \  the relation matrix $A$; the no. of elements $n$\\
	\hspace*{-20pt} \textit{Output:}  $W=A^* $\\
	1 \'  $W \leftarrow A$ \\
	2 \' \key{for} \= $k \leftarrow 1$ \key{to} $n$ \\
	3 \'           \> \key{do}  \key{for} \= $i \leftarrow 1$ \key{to} $n$ \\
	4 \'           \>                      \>  \key{do} \key{for} \= $j \leftarrow 1$ \key{to} $n$ \\ 
	5 \'            \>                     \>             \>  \key{do} \key{if} \= $w_{ik}=1$ and $w_{kj}=1$\\ 
    6 \'  \> \> \> \> \key{then} $w_{ij}\leftarrow 1$   \\ 
	7 \'  \key{return} $W$ 
\end{algN}

\begin{figure}[th]
\begin{minipage}[c]{6cm}
\includegraphics[scale=0.4]{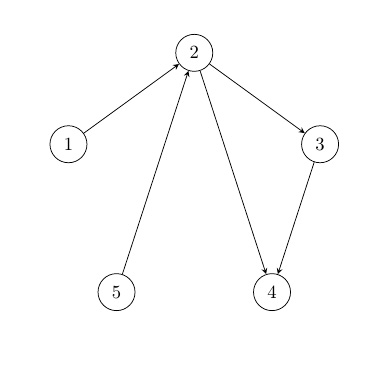}
\end{minipage}\qquad\qquad
\begin{minipage}[c]{6cm}
$\left( 
\begin{array}{ccccc}
0 &1& 0& 0& 0\\ 
0 &0& 1& 1& 0\\ 
0 &0 &0 &1 &0 \\
0 &0 &0 &0 &0 \\
0 &1 &0 &0& 0
\end{array}
\right)$
\end{minipage}
\caption{A binary relation represented by a graph with the corresponding adjacency matrix\label{fig1}}
\end{figure}

\begin{figure}[!h]
	\begin{minipage}[c]{6cm}
		\includegraphics[scale=0.45]{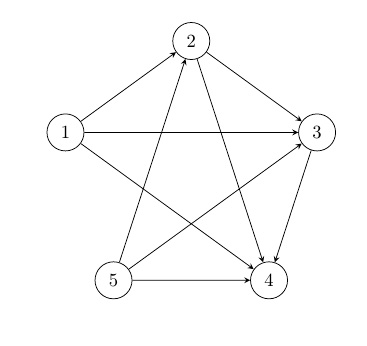}
	\end{minipage}\qquad\qquad
	\begin{minipage}[c]{6cm}
		$\left( 
		\begin{array}{ccccc}
	  0&      1&      1&      1&    0\\ 
	0 &     0&      1&      1&      0\\ 
	0 &     0 &     0 &     1 &     0 \\
	0  &    0  &    0  &    0  &    0 \\
	0   &   1   &   1   &   1   &   0 
		\end{array}
		\right)$
	\end{minipage}
	\caption{The transitive closure of the relation in Fig. \ref{fig1}\label{fig2}}
\end{figure}

\noindent A binary relation can be represented by a directed graph (i.e. digraph) too. he relation matrix is equal to the adjacency matrix of the corresponding graph. See Fig.\! \ref{fig1} for an example. Fig.\!  \ref{fig2} represents the graph of the corresponding transitive closure relation.

\newpage\section{Generalization of Warshall's algorithm}

Lines 5 and 6 in the Warshall's algorithm described above can be changed in 
 
 \[w_{ij}\leftarrow w_{ij}+ w_{ik}w_{kj}\] 
 
\noindent using the operations defined above. If instead of the operations + and $\cdot$ we  use two operations  $\oplus$ and $\odot$ from a semiring, a  generalized Warshall's algorithm results \cite{robert}: 

\begin{algN}{Generalized-Warshall($A,n$)} 
	\hspace*{-20pt} \textit{Input:} \  the relation matrix $A$; the no. of elements $n$\\
	\hspace*{-20pt} \textit{Output:}  $W=A^* $\\
	1 \'  $W \leftarrow A$ \\
	2 \' \key{for} \= $k \leftarrow 1$ \key{to} $n$ \\
	3 \'           \> \key{do}  \key{for} \= $i \leftarrow 1$ \key{to} $n$ \\
	4 \'           \>                      \>  \key{do} \key{for} \= $j \leftarrow 1$ \key{to} $n$ \\ 
	5 \'            \>                     \>             \>  \key{do} $w_{ij}\leftarrow w_{ij} \oplus (w_{ik}\odot w_{kj})$  \\ 
	6 \'  \key{return} $W$ 
\end{algN}

\noindent This generalization leads us to a number of interesting applications.

\section{Applications}

\subsection{Distances between vertices. The Floyd-Warshall's algorithm}
 
 Given a weighted (di)graph with the modified adjacency matrix $D_0=(d^0_{ij})$, we can obtain the distance matrix $D=(d_{ij})$ in which  $d_ij$ represents the distance between vertices $v_i$ and $v_j$. The distance between vertices $v_i$ and $v_j$ is the length of the shortest path between them. The modified adjacency matrix 
 $D_0=(d^0_{ij})$ is the following:
 
 \[d^0_{ij}= \left\{
 \begin{array}{ll}
 0 &  \textrm{if } i=j\\
 \infty &  \textrm{if there is no edge from vertex } v_i \textrm{ to vertex } v_j\\
 w_{ij} &  \textrm{the weight of the  edge from } v_i \textrm{ to } v_j 
 \end{array}
 \right.\]
 
\noindent Choosing for $\oplus$ the $\min$  operation (minimum between two reals), and for $\odot$ the real addition ($+$), we obtain the well-known Floyd-Warshall's algorithm as a special case of the generalized Warshall's  algorithm \cite{robert,vattai} :  
 
\begin{algN}{Floyd-Warshall($D_0,n$)}
		\hspace*{-20pt} \textit{Input:} \  the adjacency matrix $D_0$; the no. of elements $n$\\
	\hspace*{-20pt} \textit{Output:}  the distance matrix $D$\\
	1 \'  $D \leftarrow D_0$\\
	2 \' \key{for} \= $k \leftarrow 1$ \key{to} $n$ \\
	3 \'           \> \key{do}  \key{for} \= $i \leftarrow 1$ \key{to} $n$ \\
	4 \'           \>                      \>  \key{do} \key{for} \= $j \leftarrow 1$ \key{to} $n$ \\ 
	5 \'            \>                     \>             \>  \key{do} $d_{ij}\leftarrow \min \{d_{ij}, \ d_{ik}+d_{kj}\}$   \\ 
	6 \'  \key{return} $D$
\end{algN}

\begin{figure}[!h]\center
\includegraphics[scale=0.45]{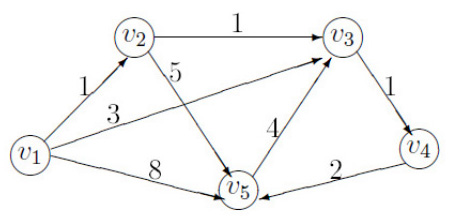}
\caption{A weighted digraph \label{floyd1}}
\end{figure}

\begin{figure}[!h]
\quad\includegraphics[scale=0.4]{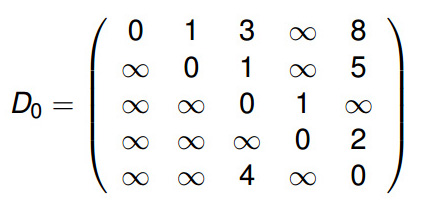}\qquad
\includegraphics[scale=0.42]{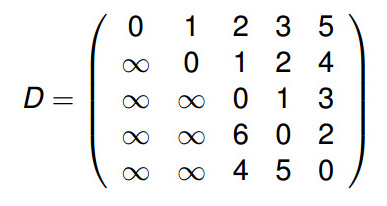}
\caption{The corresponding matrices of the Fig. \ref{floyd1}\label{floyd2}}
\end{figure}

\noindent Figures \ref{floyd1} and \ref{floyd2} contain az example.  The shortest paths can be easily obtained if 
in the description of the algorithm in line 5 we store also the previous vertex $v_k $ on the path. In the case of acyclic digraphs, the algorithm can be easily modified to obtain the longest distances between vertices, and consequently the longest paths.

\subsection{Number of paths in acyclic digraphs}
Here by path we understand directed path. In an acyclic digraph the following algorithm count the number of paths  between vertices \cite{kasa,elzinga}. The operation $\oplus, \odot$ are the classical add and multiply operations for real numbers.

\begin{algN}{Warshall-Path($A,n$)}
		\hspace*{-20pt} \textit{Input:} \  the adjacency matrix $A$; the no. of elements $n$\\
	\hspace*{-20pt} \textit{Output:}  $W$  with no. of paths between vertices\\
	1 \'  $W \leftarrow A$\\
	2 \' \key{for} \= $k \leftarrow 1$ \key{to} $n$ \\
	3 \'           \> \key{do}  \key{for} \= $i \leftarrow 1$ \key{to} $n$ \\
	4 \'           \>                      \>  \key{do} \key{for} \= $j \leftarrow 1$ \key{to} $n$ \\ 
	5 \'            \>                     \>             \>  \key{do} $w_{ij}\leftarrow w_{ij}+w_{ik}w_{kj}$   \\ 
	6 \'  \key{return} $W$
\end{algN}

\begin{figure}[!h]\center
	\includegraphics[scale=0.6]{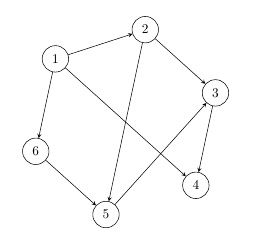}
	\caption{An acyclic digraph \label{utak}}
\end{figure}

\noindent An example can be seen in Figures \ref{utak} and \ref{utak1}. For example between vertices 1 and 3 there are 3 paths: (1,2,3); (1,2,5,3) and (1,6,5,3).

\begin{figure}[!h]\center
$A=\left(
\begin{array}{cccccc}
	0& 1& 0& 1& 0& 1\\ 
	0& 0& 1& 0& 1& 0\\ 
	0& 0& 0& 1& 0& 0 \\
	0& 0& 0& 0& 0& 0 \\
	0& 0& 1& 0& 0& 0 \\
	0& 0& 0& 0& 1& 0 
\end{array}
\right)$\qquad
$W=\left(\begin{array}{cccccc}
0&      1&      3&      4&      2&      1\\ 
0 &     0 &     2 &     2 &     1 &     0 \\
0  &    0  &    0 &     1  &    0  &    0 \\
0   &   0   &   0  &    0   &   0   &   0 \\
0    &  0    &  1   &   1    &  0    &  0 \\
0     & 0     & 1    &  1     & 1     & 0 
\end{array}
\right)$
\caption{Matrices of the example in Fig. \ref{utak}}\label{utak1}
\end{figure}

\subsection{All paths in digraphs}

\noindent The Warshall's algorithm combined with the Latin square method can be used to obtain all paths in a (not necessarily acyclic) digraph \cite{kasa}. A path will be denoted by a string formed by its vertices in there natural order in the path.

\medskip
\noindent  Let us consider a matrix ${\cal A}$ with the elements $A_{ij}$ which are set of strings. Initially  elements of this matrix are defined as:
\[A_{ij}=\left\{  \begin{array}{cl}
\{v_iv_j\},  &  \textrm{if there exists  an arc from } v_i \textrm{ to } v_j,\\
\emptyset,     &  \textrm{otherwise},
\end{array}  \quad \textrm{ for } \, i,j=1,2,\ldots, n, 
\right.
\]
If ${\cal A}$ and ${\cal B}$ are sets of strings, ${\cal AB}$ will be formed by the set of concatenation of each string from ${\cal A}$ with each string from ${\cal B}$, if they have no common elements: 
\[ {\cal AB} = \big\{ ab  \, \big| \, a\in {\cal A},  b\in {\cal B}, \textrm{ if the strings } a \textrm{ and } b \textrm{ have no common elements} \big\}.
\] 
If $s=s_1s_2\cdots s_p$ is a string, let us denote by $'s$ the string obtained from $s$ by eliminating the first character: $'s=s_2s_3\cdots s_p$.  Let us denote by $'{A_{ij}}$ the set ${A_{ij}}$ in which we eliminate from each element the first character. In this case $'{\cal A}$ is a matrix with elements $'A_{ij}.$

\medskip\noindent Operations are: the set union and set product defined as before.

\medskip\noindent  Starting with the matrix ${\cal A}$ defined as before, the algorithm to obtain all paths is the following:

\begin{algN}{Warshall-Latin(${\cal A},n$)}
		\hspace*{-20pt} \textit{Input:} \  the adjacency matrix ${\cal A }$; the no. of elements $n$\\
\hspace*{-20pt} \textit{Output:}  ${\cal W}$  matrix  of the paths between vertices\\
	1 \'  ${\cal W} \leftarrow {\cal A} $\\
	2 \' \key{for} \= $k \leftarrow 1$ \key{to} $n$ \\
	3 \'           \> \key{do}  \key{for} \= $i \leftarrow 1$ \key{to} $n$ \\
	4 \'           \>                      \>  \key{do} \key{for} \= $j \leftarrow 1$ \key{to} $n$ \\ 
	5 \'            \>                     \>             \>  \key{do} \key{if} \= $W_{ik}\ne \emptyset$ and  $W_{kj}\ne \emptyset$ \\
	6 \'        \>      \>             \>   \> \key{then} $W_{ij}\leftarrow W_{ij} \cup W_{ik}\, 'W_{kj}$   \\ 
	7 \'  \key{return} ${\cal W}$
\end{algN}

\begin{figure}[!h]\center
	\includegraphics[scale=0.5]{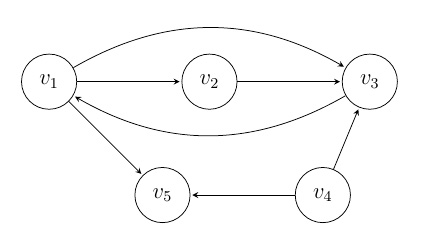}
	\caption{An example of digraph for all paths problem}\label{allpath}
\end{figure}

\noindent In Figures \ref{allpath} and \ref{allpathm} an example is given. For example between vertices $v_1 $ and $v_3 $ there are two paths: $v_1v_3$ and $v_1v_2v_3$.

\begin{figure}[!h]\center
${\cal A}=\left(
\begin{array}{ccccc}
\emptyset&\{v_1v_2\}&\{v_1v_3\}&\emptyset&\{v_1v_5\}\\
\emptyset&\emptyset&\{v_2v_3 \}&\emptyset&\emptyset\\
\{v_3v_1\}&\emptyset&\emptyset&\emptyset&\emptyset\\
\emptyset&\emptyset&\{v_4v_3\}&\emptyset&\{v_4v_5\}\\
\emptyset&\emptyset&\emptyset&\ \ \emptyset\ \ &\emptyset
\end{array}
 \right)	
$	

\bigskip
${\cal W}=\left(
	\begin{array}{ccccc}
	\emptyset&\{v_1v_2\}&\{v_1v_3, \,v_1v_2v_3\}&\emptyset&\{v_1v_5\}\\
	\{v_2v_3v_1\}&\emptyset&\{v_2v_3 \}&\emptyset&\{v_2v_3v_1v_5\}\\
	\{v_3v_1\}&\{v_3v_1v_2\}&\emptyset&\emptyset&\{v_3v_1v_5\}\\
	\{v_4v_3v_1\}&\{v_4v_3v_1v_2\}&\{v_4v_3\}&\emptyset&\{v_4v_5\}\\
	\emptyset&\emptyset&\emptyset&\ \ \emptyset\ \ &\emptyset
	\end{array}
	\right)	
	$	
\caption{Matrices for graph in Fig. \ref{allpath}}\label{allpathm}
\end{figure}

\newpage
\subsection{Scattered complexity for rainbow words}

The application mentioned here can be found in  \cite{kasa}. 
Let $\Sigma$ be an alphabet, $\Sigma^n$ the set of all length-$n$ words over $\Sigma$, $\Sigma^*$ the set of all finite word over $\Sigma$.

\begin{definition}
	Let  $n$ and $s$ be positive integers, $M\subseteq \{1,2,\ldots, n-1\}$ and $u=x_1x_2\ldots x_n\in \Sigma^n$. An \textbf{$M$-subword} of length $s$ of $u$ is defined as $v=x_{i_1}x_{i_2}\ldots x_{i_s}$ where 
	
	$i_1\ge 1$, 
	
	$i_{j+1}-i_j\in M$  for $j=1,2,\ldots, s-1$,
	
	$i_s\le n.$
\end{definition}

\begin{definition}
	The number of $M$-subwords of a word $u$ for a given set $M$ is the scattered subword complexity, simply $M$-complexity.
\end{definition}

\noindent\textbf{Examples.}  The word $abcd$ has 11 $\{1,3\}$-subwords: $a$,  $ab$, $abc$, $abcd$, $ad$, $b$, $bc$, $bcd$, $c$, $cd$,  $d$.  The $\{2,3\,4,5\}$-subwords of the word $abcdef$  are the following: $a$, $ac$, $ad$, $ae$, $af$, $ace$, $acf$, $adf$, $b$, $bd$, $be$, $bf$, $bdf$, $c$, $ce$, $cf$, $d$,  $df$, $e$, $f$.

\medskip\noindent Words with different letters are called \emph{rainbow words}. 
The $M$-complexity of a length-$n$  rainbow word does not depend on what letters it contains, and is denoted by $K(n,M)$.

\medskip\noindent To compute the $M$-complexity of a rainbow word of length $n$ we will use graph theoretical results. 
 Let us consider the rainbow word  $a_1a_2\ldots a_n$ and the corresponding digraph $G=(V, E)$, with 

$V=\big\{ a_1, a_2, \ldots, a_n \big\}$, 

$E=\big\{ (a_i,a_j) \mid j-i\in M, \, i=1,2,\ldots, n,  j=1,2,\ldots, n  \big\}$. 

\begin{figure}
\begin{center}
\begin{tikzpicture}[scale=2]
\begin{scope}[>=latex]
\filldraw[]     (1,2) circle (2pt) 
                (2,2) circle (2pt)
                (3,2) circle (2pt)  
                (4,2) circle (2pt)
                (5,2) circle (2pt) 
                (6,2) circle (2pt);
\draw [->] (1,2) .. controls (1.5,2.5) and (2.5,2.5) .. (2.95,2.05);
\draw [->] (1,2) .. controls (2,2.8) and (3,2.8) .. (3.95,2.05);
\draw [->] (1,2) .. controls (2,3.1) and (4,3.1) .. (4.95,2.05);
\draw [->] (1,2) .. controls (2,3.4) and (5,3.4) .. (5.95,2.05);
\draw [->] (3,2) .. controls (3.5,2.5) and (4.5,2.5) .. (4.95,2.05);
\draw [->] (3,2) .. controls (4,2.8) and (5,2.8) .. (5.95,2.05);
\draw [->] (4,2) .. controls (4.5,1.4) and (5.5,1.4) .. (5.95,1.95);
\draw [->] (2,2) .. controls (2.5,1.4) and (3.5,1.4) .. (3.95,1.95);
\draw [->] (2,2) .. controls (3,0.9) and (4,0.9) .. (4.95,1.95);
\draw [->] (2,2) .. controls (2.5,0.6) and (5.5,0.6) .. (5.95,1.95);
\node (1) at (1,1.75) {$a$};
\node (2) at (1.95,1.75) {$b$};
\node (3) at (3,1.75) {$c$};
\node (4) at (4,1.75) {$d$};
\node (5) at (5,1.75) {$e$};
\node (6) at (6.05,1.75) {$f$};
\end{scope}
\end{tikzpicture}
\end{center}\vspace*{-0.8cm}
\caption{Graph for $\{2,3,4,5\}$-subwords of the rainbow word of  length-6 \label{fig9}}
\end{figure}
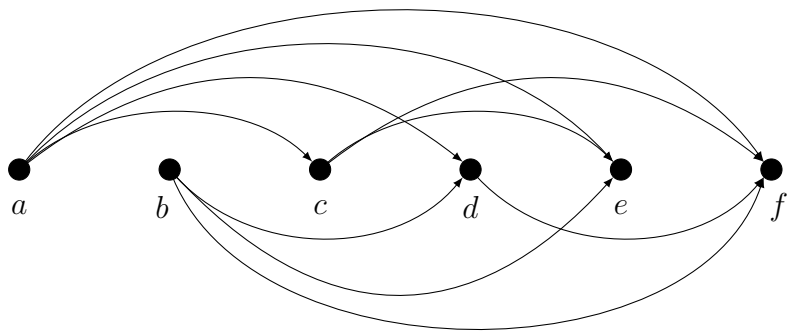

\medskip\noindent
For $n=6, M=\{2,3,4,5\}$ see Fig. \ref{fig9}.

\medskip\noindent The adjacency matrix  $A=\big(a_{ij}\big)_{\!\!\!\tiny\begin{array}{c}i\!\!=\!\!\overline{1,\!n}\\
                                                            j\!\!=\!\!\overline{1,\!n}\end{array}}$ 
of the graph is defined by:

\[a_{ij}=\left\{  \begin{array}{ll}
                    1,  &  \textrm{if } j-i\in M,\\
                    0,     &  \textrm{otherwise},
                 \end{array}  \quad \textrm{  for  }   i=1,2,\ldots, n, j=1,2,\ldots, n.                        
           \right.
\]

\medskip\noindent Because the graph has no directed cycles, the element in row $i$ and column $j$ in  $A^k$ (where $A^k=A^{k-1}A$, with $A^1=A$) will represent the number of length-$k$  directed paths from $a_i$ to $a_j$.  If $I$ is the identity matrix  (with elements equal to 1 only on the first diagonal, and 0 otherwise), let us define the matrix $R= (r_{ij})$:
\[R= I+A+A^2+\cdots + A^k, \textrm{ where } A^{k+1}=O \, (\textrm{the null matrix}).\]
The $M$-complexity of a rainbow word is then
\[ K(n,M)= \sum_{i=1}^{n}{\sum_{j=1}^{n}{r_{ij}}}.\]
Matrix $R$  can be better computed using the \textsc{Warshall-Path} algorithm.  

\medskip\noindent From $W$ we obtain easily $R=I+W$.

\noindent For example let us consider the graph in Fig. \ref{fig9}. The corresponding adjacency matrix is:

\[A=\left(\begin{array}{cccccc}
0&0&1&1&1&1 \\
0&0&0&1&1&1 \\
0&0&0&0&1&1 \\
0&0&0&0&0&1 \\
0&0&0&0&0&0 \\
0&0&0&0&0&0 \\
\end{array}
\right) 
\]

\noindent After applying the \textsc{Warshall-Path}  algorithm:
\[W=\left(\begin{array}{cccccc}
0&0&1&1&2&3 \\
0&0&0&1&1&2 \\
0&0&0&0&1&1 \\
0&0&0&0&0&1 \\
0&0&0&0&0&0 \\
0&0&0&0&0&0 \\
\end{array}
\right),
\qquad 
R=\left(\begin{array}{cccccc}
1&0&1&1&2&3 \\
0&1&0&1&1&2 \\
0&0&1&0&1&1 \\
0&0&0&1&0&1 \\
0&0&0&0&1&0 \\
0&0&0&0&0&1 \\
\end{array}
\right)
\]
and then $K\big(6,\{2,3,4,5\}\big)=20,$  the sum of elements in $R$.

\medskip\noindent Using the \textsc{Warshall-Latin} algorithm we can  obtain all nontrivial (with length at least 2) $M$-subwords of a given length-$n$ rainbow word $a_1a_2\cdots a_n$. Let us consider a matrix ${\cal A}$ with the elements $A_{ij}$ which are set of strings. Initially this matrix is defined as:
\[A_{ij}=\left\{  \begin{array}{ll}
                    \{a_ia_j\},  &  \textrm{if } j-i\in M,\\
                    \emptyset,     &  \textrm{otherwise},
                 \end{array}  \quad \textrm{ for } \, i=1,2,\ldots, n, \, j=1,2,\ldots, n. 
           \right.
\]

\medskip\noindent The set of nontrivial $M$-subwords is ${\displaystyle\bigcup_{i,j\in \{ 1,2,\ldots, n \} } W_{ij}}$. 

\medskip\noindent For $n=8$, $M=\{3,4,5,6,7\}$ the initial matrix is:
\begin{center}
$\left(\begin{array}{cccccccc}
\emptyset & \emptyset & \emptyset & \{ad\} & \{ae\} &\{af\} &\{ag\} &\{ah\} \\ 
\emptyset & \emptyset & \emptyset & \emptyset & \{be\}& \{bf\} & \{bg\}& \{bh\} \\ 
\emptyset & \emptyset & \emptyset & \emptyset & \emptyset & \{cf\}&  \{cg\} & \{ch\} \\ 
\emptyset & \emptyset & \emptyset & \emptyset & \emptyset &\emptyset & \{dg\} &\{dh\} \\ 
\emptyset & \emptyset & \emptyset & \emptyset & \emptyset &\emptyset &\emptyset & \{eh\}\\ 
\emptyset & \emptyset & \emptyset & \emptyset & \emptyset &\emptyset &\emptyset & \emptyset \\
\emptyset & \emptyset & \emptyset & \emptyset & \emptyset &\emptyset &\emptyset & \emptyset \\
\emptyset & \emptyset & \emptyset & \emptyset & \emptyset &\emptyset &\emptyset & \emptyset \\
\end{array}
\right).
$
\end{center}

\medskip\noindent The result of the algorithm in this case is: 

\begin{center}
$\left(\begin{array}{cccccccc}
\emptyset & \emptyset & \emptyset & \{ad\} & \{ae\} &\{af\} &\{ag, adg\} &\{ah,adh,aeh\} \\ 
\emptyset & \emptyset & \emptyset & \emptyset & \{be\}& \{bf\} & \{bg\}& \{bh,beh\} \\ 
\emptyset & \emptyset & \emptyset & \emptyset & \emptyset & \{cf\}&  \{cg\} & \{ch\} \\ 
\emptyset & \emptyset & \emptyset & \emptyset & \emptyset &\emptyset & \{dg\} &\{dh\} \\ 
\emptyset & \emptyset & \emptyset & \emptyset & \emptyset &\emptyset &\emptyset & \{eh\}\\ 
\emptyset & \emptyset & \emptyset & \emptyset & \emptyset &\emptyset &\emptyset & \emptyset \\
\emptyset & \emptyset & \emptyset & \emptyset & \emptyset &\emptyset &\emptyset & \emptyset \\
\emptyset & \emptyset & \emptyset & \emptyset & \emptyset &\emptyset &\emptyset & \emptyset \\
\end{array}
\right).
$
\end{center}

\subsection{Special paths in finite automata}

Let us consider a finite automaton 
$A=(Q, \Sigma,\delta, \{q_0\}, F)$, where
$Q$ is a finite set of states, $ \Sigma$  
 the input alphabet, $\delta: Q\times \Sigma \rightarrow Q$  the transition function, $q_0$ the initial state,  $F$ the set of finale states. In following we do not need to mark the initial and the finite states.

\medskip\noindent The transition function can be generalized for words too: $\delta(q,wa)=\delta(\delta(q,w), a)$, where $q\in Q, a \in \Sigma,w\in\Sigma^*$.

\medskip\noindent We are interesting in finding for each pair $p, q$ of states the letters $a$ for which there exists a natural $k\ge 1$ such that  we have the transition $\delta(p,a^k) =q$ \cite{robert}, i.e.:
\[ w_{pq}=\{a\in \Sigma \mid \exists k\ge 1,    \delta(p,a^k) =q\}.\]

\bigskip
\noindent Instead of $\oplus$ we use here set union  ($\cup$) and instead of $\odot$  set intersection ($\cap$).

\begin{algN}{Warshall-Automata($A,n$)}
	\hspace*{-20pt} \textit{Input:} \  the adjacency matrix $A$; the no. of elements $n$\\
	\hspace*{-20pt} \textit{Output:}  $W$  with sets of states\\
	1 \'  $W \leftarrow A$\\
	2 \' \key{for} \= $k \leftarrow 1$ \key{to} $n$ \\
	3 \'           \> \key{do}  \key{for} \= $i \leftarrow 1$ \key{to} $n$ \\
	4 \'           \>                      \>  \key{do} \key{for} \= $j \leftarrow 1$ \key{to} $n$ \\ 
	5 \'            \>                     \>             \>  \key{do} $w_{ij}\leftarrow w_{ij}\cup (w_{ik}\cap w_{kj})$   \\ 
	6 \'  \key{return} $W$
\end{algN}

\begin{figure}[!h]
\includegraphics[scale=0.7]{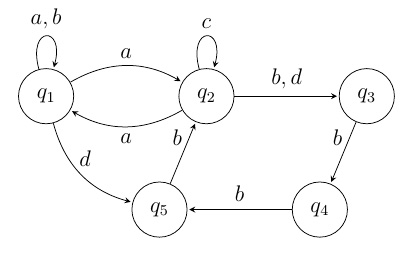}
\caption{An example of a finite automaton without indicating the initial and finite states\label{aut1}}
\end{figure}

\newpage
\noindent The transition table of the finite automaton in Fig. \ref{aut1} is:

\bigskip\qquad
$
\begin{array}{c|cccc}
 \delta &a &b& c&d \\ \hline
q_1 &\{q_1,q_2\} &\{q_1\} & \emptyset& \{q_5\}\\
q_2 &  \{q_1\}& \{q_3\}&  \{q_2\}&\{q_3\} \\
q_3 &  \emptyset& \{q_4\}&  \emptyset&  \emptyset\\
q_4 &  \emptyset&  \{q_5\}&  \emptyset& \emptyset\\
q_5 & \emptyset & \{q_2\}&  \emptyset&  \emptyset
\end{array}
$

\bigskip\bigskip\noindent Matrices for  graph in Fig. \ref{aut1} are the following:

\bigskip\bigskip
 $
A=\left(\begin{array}{ccccc}
\{a,b\}&\{a\} &\emptyset& \emptyset&\{d\} \\ 
 \{a\}& \{c\}& \{b,d\}&\emptyset &\emptyset \\
\emptyset& \emptyset&\emptyset &\{b\} &\emptyset \\
 \emptyset&\emptyset &\emptyset  &\emptyset &\{b\} \\
\emptyset & \{b\} &\emptyset&\emptyset &\emptyset 
\end{array}
\right)
$

\bigskip\bigskip
$
W=\left(\begin{array}{ccccc}
\{a,b\}&\{a\} &\emptyset& \emptyset&\{d\} \\ 
\{a\}& \{a,b,c\}& \{b,d\}&\{b\} &\{b\} \\
\emptyset& \{b\}&\{b\} &\{b\} &\{b\} \\
\emptyset& \{b\}&\{b\} &\{b\} &\{b\} \\
\emptyset& \{b\}&\{b\} &\{b\} &\{b\} \\
\end{array}
\right)
$

\bigskip\medskip\noindent  For example $\delta(q_2,bb)=q_4$,  
$\delta(q_2,bbb)=q_5$,
$\delta(q_2,bbbb)=q_2$, $\delta(q_2,c^k)=q_2$ for $k\ge 1$.


\begin{thebibliography}{99}

\bibitem{baase} S. {Baase}, \emph{Computer Algorithms: Introduction to Design and Analysis,} Second edition, Addison--Wesley, 1988.

\bibitem{elzinga} C. \href{https://scholar.google.nl/citations?user=0feufd8AAAAJ&hl=nl}{Elzinga},  H. Wang, \href{https://www.sciencedirect.com/science/article/pii/S030439751300460X?via\%3Dihub}{Versatile string kernels},   \emph{Theor. Comput. Sci.}, \textbf{495,} 1 (2013)  50--65.


\bibitem{kasa} Z. \href{http://www.ms.sapientia.ro/~kasa}{K\'asa}, On \href{http://acta.sapientia.ro/acta-info/C3-1/info31-6.pdf}{scattered subword complexity} 
\href{Acta Universitatis Sapientiae, Informatica,} \textbf{3,} 1 (2011) 127--136.  



\bibitem{robert} P. Robert, J. Ferland, G\'en\'eralisation de l'algorithme de Warshall, \textit{Revue Fran\c caise d'Informatique et de Recherche Op\'erationnelle} \textbf{2,} 7 (1968) 71--85. \url{http://www.numdam.org/item/?id=M2AN_1968__2_1_71_0}


\bibitem{vattai} Z. A. \href{http://www.ekt.bme.hu/Szemallo/Gbvattai.shtml}{Vattai}: Floyd-Warshall again, \url{http://www.ekt.bme.hu/Cikkek/54-Vattai_Floyd-Warshall_Again.pdf}

\bibitem{warshall} S. \href{https://en.wikipedia.org/wiki/Stephen_Warshall}{Warshall}, A \href{http://bioinfo.ict.ac.cn/~dbu/AlgorithmCourses/Lectures/Warshall1962.pdf}{ theorem on boolean matrices}, {\it Journal of the ACM}, \textbf{9,} 1 (1962) 11--12.
\end{thebibliography}
\end{document}